\documentclass[10pt,a4paper]{article}
\usepackage[T1]{fontenc}
\usepackage{amsmath,amsfonts, amssymb, amsbsy, mathrsfs, mathtools, bbm, enumitem}
\usepackage{titlesec}
\usepackage{float}
\usepackage{makeidx}
\usepackage{graphicx}
\usepackage{color}
\usepackage{url}
\usepackage{authblk}
\usepackage{multirow}
\usepackage{lscape}

\usepackage{mathpazo}
\usepackage[toc]{appendix}
\usepackage[hidelinks,colorlinks=true,linkcolor=red,citecolor=blue,urlcolor=magenta]{hyperref}
\usepackage[left=2.7cm, right=2.7cm, top=3.00cm, bottom=3.00cm]{geometry}

\usepackage{fancyhdr}

\newcommand{\email}[1]{\href{mailto:#1}{#1}}

\newcommand{\df}{\textrm{d}}
\newcommand{\Df}{\textrm{D}}
\newcommand{\cg}{\textnormal{\textsl{g}}}

\newcommand{\e}{\textrm{e}}
\newcommand{\hf}{{\frac{1}{2}}}

\newcommand{\tr}{{^{{\rm (T)}}}}

\renewcommand{\bar}{\overline}

\titleformat{\paragraph}
{\normalfont\normalsize\bfseries}{\theparagraph}{1em}{}

\titleformat{\subparagraph}
{\normalfont\normalsize\bfseries}{\thesubparagraph}{1em}{}

\usepackage[square,numbers,sort&compress]{natbib}

\numberwithin{equation}{section}

\allowdisplaybreaks[1]

\begin{document}
	\setlength{\bibsep}{0pt}
	
	\title{ \textbf{Lagrangian Formulation of the Raychaudhuri Equation in Non-Riemannian Geometry}}
	\author[]{Anish Agashe\thanks{\email{anagashe@smcm.edu}}}
	\affil[]{\small Department of Physics, St. Mary's College of Maryland,\\ 47645 College Dr, St. Mary's City, MD 20686 USA}
	\date{\small \today}

	\maketitle
	
	\begin{abstract}
		The Raychaudhuri equation for a congruence of curves in a general non-Riemannian geometry is derived. A formal connection is established between the expansion scalar and the cross-sectional volume of the congruence. It is found that the expansion scalar is equal to the fractional rate of change of volume, weighted by a scalar factor that depends on the non-Riemannian features of the geometry. Treating the congruence of curves as a dynamical system, an appropriate Lagrangian is derived such that the corresponding Euler-Lagrange equation is the Raychaudhuri equation. A Hamiltonian formulation and Poisson brackets are also presented.
	\end{abstract}

	\section{Introduction}	
	Given a geometry, two independent features can be endowed to it, namely, metric and connection. A metric formalises the notion of distances, while a connection formalises the notion of parallelism \cite{schout}. General relativity (GR) employs a geometrical framework in which the connection can be expressed solely in terms of the metric and its (partial) derivatives. This connection is both symmetric and metric compatible and is commonly known as Levi-Civita connection and such a geometry is called Riemannian. On the other hand, geometries in which any one of the aforementioned conditions (symmetry or metric-compatibility) is violated are called non-Riemannian. The focus of this paper is a general non-Riemannian geometry, where both these conditions are violated.
	
	Using a non-Riemannian geometrical framework, several extensions of GR have been proposed \cite{weyl1,cart0,cart1,cart2,cart3,cart4,jim,capo4,haya,maluf,aldro,baha,ferr,nest,capo1,hehl1}. With an increasing interest in these theories, the mean kinematics of congruence of curves in their underlying geometries have also been analysed \cite{luz,hensh,dey1,dey2,dey3,capo2,wanas,cai,pasma,spez,lobo,iosth,ios,yang,aga}. Given a congruence of curves (timelike or null), its kinematics can be described in terms of the irreducible components of the (transverse part of the) gradient of the tangent to the congruence. These components are known as expansion (the trace part), rotation (the antisymmetric part), and shear (the symmetric traceless part). Then, one can derive equations governing the evolution of these variables purely in terms of geometrical quantities. The equation for the evolution of expansion of a congruence of timelike curves has been historically known as the Raychaudhuri equation\footnote{Strictly speaking, this is a geometrical identity rather than an equation. However, it can be converted into an equation by using it in conjunction with the field equations of a given gravity theory.} \cite{ray1}.
	
	In this paper, we present a Lagrangian (and a Hamiltonian) formulation of the Raychaudhuri equation (RE). This has been previously done in Riemannian geometry \cite{chakra,horw,alsa}. Here, we extend this to a general non-Riemannian geometry. To be able to do this, the RE needs to be written as a second-order differential equation. In Riemannian geometry, this is done by the virtue of the fact that the expansion scalar is equal to the fractional rate of change of the cross-sectional volume of the congruence \cite{alsa,chakra,horw}. However, we find that not to be the case in a non-Riemannian geometry. Instead, the expansion scalar is related to the fractional rate of change of a scalar variable that is a combination of the cross-sectional volume and the non-Riemannian features of the geometry. Using this variable as the dynamical degree of freedom (dof), we construct a Lagrangian, the corresponding Euler-Lagrange equation of which, is the RE. We also provide a Hamiltonian formulation and Poisson brackets for the RE.
	
	The paper is arranged in the following manner: in section \ref{sec-nrgbasics}, we introduce basics of non-Riemannian geometry. Then, in section \ref{sec-nrgkin}, we follow steps analogous to the Riemannian geometry to derive the generalised RE. We present the Lagrangian and Hamiltonian formulation of the Raychaudhuri equation in section \ref{sec-raylagham}. Finally, in section \ref{sec-discuss}, we discuss the results and provide concluding remarks.
	
	The notation and convention used is as follows: Greek indices run from $0$  to $n-1$ and Latin indices from $1$ to $n-1$. Indices with round brackets $ (~) $/square brackets $[~]$ are symmetrised/anti-symmetrised; and  underlined indices are not included in (anti-)symmetrisation. The partial derivative is denoted by $ \boldsymbol{\partial} $; covariant derivative is denoted by $\boldsymbol{\nabla}$; the directional derivative along a vector, $ \boldsymbol{X} $, is denoted by $ \boldsymbol{\Df_X} \equiv \boldsymbol{X\cdot\nabla} $; and the Lie derivative of a vector, $ \boldsymbol{Y} $, along another vector, $ \boldsymbol{X} $, denoted by $ \boldsymbol{\mathfrak{L}_X Y} \equiv \boldsymbol{X\cdot\partial Y - Y\cdot\partial X} $. The sign convention followed is the `Landau-Lifshitz Space-like Convention (LLSC)' \cite{mtw}.
	
	\section{Basics of non-Riemannian Geometry} \label{sec-nrgbasics}
	Consider an arbitrary $ n $-dimensional non-Riemannian manifold with a metric, $ \boldsymbol{g} $, and a connection, $ \boldsymbol{\Gamma} $. The connection is neither metric compatible nor symmetric. These features of the connection are used to define two quantities -- torsion and non-metricity -- that take into account the non-Riemannian nature of the geometry. The  torsion, $ \boldsymbol{T} $, is defined as the antisymmetric part of the connection,
	\begin{equation}\label{torsion}
		{T^\rho}_{\alpha\beta} = {\Gamma^\rho}_{[\alpha\beta]} = \hf \left( {\Gamma^\rho}_{\alpha\beta} - {\Gamma^\rho}_{\beta\alpha}\right)
	\end{equation}
	and non-metricity, $ \boldsymbol{Q} $, is defined as\footnote{The covariant derivative is defined as,
		\begin{multline}\label{covder2}
			{\nabla}_\rho {V^{\alpha_1 \alpha_2  \dots}}_{\beta_1 \beta_2 \dots} = \partial_\rho {V^{\alpha_1 \alpha_2  \dots}}_{\beta_1 \beta_2 \dots} + {{\Gamma}^{\alpha_1}}_{\sigma\rho}{V^{\sigma \alpha_2  \dots}}_{\beta_1 \beta_2 \dots} + {{\Gamma}^{\alpha_2}}_{\sigma\rho}{V^{ \alpha_1 \sigma  \dots}}_{\beta_1 \beta_2 \dots} + \dots \\- {{\Gamma}^\sigma}_{\beta_1\rho}{V^{ \alpha_1 \alpha_2  \dots}}_{\sigma \beta_2 \dots} - {{\Gamma}^\sigma}_{\beta_2\rho}{V^{ \alpha_1 \alpha_2  \dots}}_{\beta_1 \sigma \dots} - \dots
	\end{multline}},
	\begin{equation}\label{nonmetricity}
		Q_{\alpha\beta\sigma} = \hf \nabla_\alpha \cg_{\beta\sigma} = \hf \partial_\alpha \cg_{\beta\sigma} - \hf {\Gamma^\rho}_{\beta\alpha} \cg_{\rho\sigma} - \hf {\Gamma^\rho}_{\sigma\alpha} \cg_{\beta\rho}
	\end{equation}
	
	The connection can be decomposed in the following manner,
	\begin{equation}\label{distor1}
		{\Gamma^{\rho}}_{\alpha\beta} = {\bar{\Gamma}^\rho}_{\alpha\beta} + {N^{\rho}}_{\alpha\beta} 
	\end{equation}
	where, $ \boldsymbol{\bar{\Gamma}} $ is the Riemannian part (Levi-Civita connection), and $ \boldsymbol{N} $ is the so-called distortion tensor that encapsulates the deviation from the Riemannian nature of geometry. The distortion is given in terms of the torsion and non-metricity as,
	\begin{equation}\label{distor2}
		{N^{\rho}}_{\alpha\beta} := \left( {T^{\rho}}_{\alpha\beta} - {T_\alpha}^\rho\! _\beta  - {T_\beta}^\rho\! _\alpha \right) + \left( {Q^{\rho}}_{\alpha\beta} - {Q_\alpha}^\rho\! _\beta  - {Q_\beta}^\rho\! _\alpha \right)
	\end{equation}
	The combination of the torsion terms above is called the contorsion tensor, $ \boldsymbol{K} $, given by,
	\begin{equation}\label{contor}
		{K^{\rho}}_{\alpha\beta} = {T^{\rho}}_{\alpha\beta} - {T_\alpha}^\rho\! _\beta  - {T_\beta}^\rho\! _\alpha
	\end{equation}
	and the combination of the non-metricity terms is called deformation (or disformation) tensor, $ \boldsymbol{P} $, given by,
	\begin{equation}\label{qcomb}
		{P^{\rho}}_{\alpha\beta} =  {Q^{\rho}}_{\alpha\beta} - {Q_\beta}^\rho\! _\alpha - {Q_{\alpha\beta}}^\rho
	\end{equation}
	
	The Ricci identity is given by,
	\begin{equation}\label{ricciidenma}
		\left({\nabla}_\alpha{\nabla}_\beta - {\nabla}_\beta{\nabla}_\alpha \right)V^\rho = {{R}^\rho}_{\sigma\alpha\beta}V^\sigma + 2 {T^\sigma}_{\alpha\beta} \nabla_\sigma V^\rho 
	\end{equation}
	where, $ {{R}^\rho}_{\sigma\alpha\beta} $ is the Riemann curvature tensor defined as,
	\begin{equation}\label{riemcurvma}
		{{R}^\rho}_{\sigma\alpha\beta} = \partial_\alpha {{\Gamma}^\rho}_{\sigma\beta} - \partial_\beta {{\Gamma}^\rho}_{\sigma\alpha} + {{\Gamma}^\rho}_{\mu\alpha}{{\Gamma}^\mu}_{\sigma\beta} - {{\Gamma}^\rho}_{\mu\beta}{{\Gamma}^\mu}_{\sigma\alpha}
	\end{equation}
	The Ricci tensor and Ricci scalar are defined in the usual manner as the traces of the Riemann curvature tensor\footnote{Besides the usual Ricci tensor above, there exist two more independent traces of the Riemann tensor,
		\begin{align}
			{R}_{\alpha\beta} &= \cg^{\rho\sigma}{R}_{\rho\sigma\alpha\beta} =  {{R}^\sigma}_{\sigma\alpha\beta} \label{riccitenma2}\\
			{R}_{\rho\beta} &= \cg^{\sigma\alpha}{R}_{\rho\sigma\alpha\beta} =  {{R_\rho}^\alpha}\ \!_{\alpha\beta} \label{riccitenma3}
		\end{align}
		These are called the homothetic tensor and co-Ricci tensor, respectively. Although there are three independent traces of the Riemann tensor, the Ricci scalar is uniquely defined since the trace of the homothetic tensor vanishes and that of the co-Ricci tensor is simply $ -R $.},
	\begin{align}
		{R}_{\sigma\beta} &= \cg^{\rho\alpha}{R}_{\rho\sigma\alpha\beta} =  {{R}^\alpha}_{\sigma\alpha\beta} \label{riccitenma}\\
		{R} &= \cg^{\sigma\beta}{R}_{\sigma\beta} = {{R}^{\alpha\beta}}_{\alpha\beta} \label{ricciscalarma}
	\end{align}

	\section{Raychaudhuri Equation} \label{sec-nrgkin}
	To derive the Raychaudhuri equation, we begin with considering a congruence of timelike curves. Let these curves be parametrised by a parameter, $ t $. Then, the vector field, $ u^\alpha = \frac{\partial x^\alpha}{\partial t} $, is tangent to the curves.  One of the effects of the non-metricity of the geometry is that the length of vectors is not preserved under parallel transportation \cite{ios}. Therefore, the norm of $ u^\alpha $ cannot be normalised like it is done in a Riemannian geometry. Instead, now we have,
	\begin{equation}\label{vecnormma}
		u^\alpha u_\alpha = - \ell^2
	\end{equation}
	where, $ \ell \equiv \ell(x^\alpha) $. 
	
	The Raychaudhuri equation is simply the evolution equation for the trace of the transverse component of the $n$-velocity gradient ($\nabla_\alpha u_\beta$) of the curves. To find the transverse component, we define a projection operator, given by,
	\begin{equation}\label{transmetricma}
		h_{\alpha\beta} = \cg_{\alpha\beta} + \frac{1}{\ell^2}u_\alpha u_\beta \;\; \Rightarrow\ {h^\alpha}_\beta = \delta^\alpha_\beta + \frac{1}{\ell^2}u^\alpha u_\beta
	\end{equation} 
	It is easy to check that, $ u^\alpha h_{\alpha\beta} = 0 = h_{\alpha\beta}u^\beta $; $ {h^\alpha}_\sigma {h^\sigma}_\beta = {h^\alpha}_\beta $; and $ h^{\alpha\beta}h_{\alpha\beta} = {h^\alpha}_\alpha = n-1 $.	The transverse component of the velocity gradient is then given by,
	\begin{equation}\label{transvelgrad}
		\tr \nabla_\alpha u_\beta = {h^\rho}_\alpha {h^\epsilon}_\beta \nabla_\rho u_\epsilon = \nabla_\alpha u_\beta + X_{\alpha\beta}
	\end{equation}
	where, we have defined,
	\begin{equation}\label{graddiff}
		X_{\alpha\beta} := \frac{1}{\ell^2}u^\epsilon \nabla_\alpha u_\epsilon u_\beta + \frac{1}{\ell^2}u_\alpha a_\beta + \frac{1}{\ell^4}u^\epsilon a_\epsilon u_\alpha u_\beta
	\end{equation}
	with, $a_\beta := \Df_u u_\beta = u^\alpha \nabla_\alpha u_\beta$.
	
	The kinematic variables are defined as the trace (expansion), antisymmetric (rotation), and symmetric traceless (shear) parts of the velocity gradient,
	\begin{align} \label{kinvardef}
		\theta &:= \cg^{\alpha\beta} {h^\rho}_\alpha {h^\epsilon}_\beta \nabla_\rho u_\epsilon = h^{\alpha\beta}\nabla_\alpha u_\beta\\
		\omega_{\alpha\beta} &:= {h^\rho}_{[\alpha} {h^\epsilon}_{\beta]} \nabla_\rho u_\epsilon\\ \sigma_{\alpha\beta} &:= {h^\rho}_{(\alpha} {h^\epsilon}_{\beta)} \nabla_\rho u_\epsilon - \frac{1}{n-1}\theta h_{\alpha\beta}
	\end{align}
	Using equations \eqref{transvelgrad} and \eqref{kinvardef}, we can write,
	\begin{equation}\label{velgrad}
		\nabla_\alpha u_\beta = \frac{1}{n-1}\theta h_{\alpha\beta} + \omega_{\alpha\beta} + \sigma_{\alpha\beta} - X_{\alpha\beta}
	\end{equation}
	
	The first step in deriving the Raychaudhuri equation is to write (using the Ricci identity(Eq. \eqref{ricciidenma}) the following equation,
	\begin{equation} \label{riccicontra}
		\left({\nabla}_\alpha{\nabla}_\beta - {\nabla}_\beta{\nabla}_\alpha \right)u_\rho h^{\beta\rho}u^\alpha = -R_{\alpha\beta} u^\alpha u^\beta + 2 T^{\rho\alpha\beta} u_\alpha \nabla_\rho u_\beta + 4 T^{\rho\alpha\beta} Q_{\rho\alpha\epsilon}u_\beta u^\epsilon
	\end{equation}
	Further, a straightforward calculation leads to,
	\begin{equation} \label{riccilhs1}
		\nabla_\alpha \nabla_\beta u_\rho h^{\beta\rho}u^\alpha = \Df_u \theta + \left(2{Q_\alpha}^{\beta\rho}u^\alpha - \frac{1}{\ell^2}A^\beta u^\rho\right)\nabla_\beta u_\rho - \frac{1}{\ell^2}A^\rho a_\rho 
	\end{equation}
	\begin{multline} \label{riccilhs2}
		\nabla_\beta\nabla_\alpha u_\rho h^{\beta\rho}u^\alpha = \nabla_\rho a^\rho + 2{Q^{\rho\alpha}}_\rho a_\alpha + \frac{1}{\ell^2}u^\rho \Df_u a_\rho - \cg^{\rho\alpha}\cg^{\epsilon\beta}\left(\nabla_\alpha u_\beta\right)\left(\nabla_\epsilon u_\rho\right) \\ + \left(2Q^{\rho\alpha\beta}u_\alpha + \frac{2}{\ell^2}{Q_\epsilon}^{\alpha\beta} u_\alpha u^\epsilon u^\rho - \frac{1}{\ell^2}a^\beta u^\rho\right)\nabla_\beta u_\rho
	\end{multline}
	where, $A^\beta:= \Df_u u^\beta $ and $a^\beta = \cg^{\alpha\beta}a_\alpha \ne A^\beta$. Further, one can also show that,
	\begin{equation}
		\cg^{\rho\alpha}\cg^{\epsilon\beta}\left(\nabla_\alpha u_\beta\right)\left(\nabla_\epsilon u_\rho\right) = \frac{1}{n-1}\theta^2 - \omega^2 + \sigma^2 + X^2
	\end{equation}
	where, $\sigma^2 = \cg^{\rho\alpha}\cg^{\epsilon\beta}\sigma_{\alpha\beta}\sigma_{\rho\epsilon}$; $\omega^2 = \cg^{\rho\alpha}\cg^{\epsilon\beta}\omega_{\alpha\beta}\omega_{\rho\epsilon}$; and $X^2 = \cg^{\rho\alpha}\cg^{\epsilon\beta}X_{\alpha\beta}X_{\epsilon\rho}$.
	
	Using equations \eqref{riccilhs1} and \eqref{riccilhs2} in Eq. \eqref{riccicontra}, we get,
	\begin{multline}
		\Df_u \theta + \left(2{Q_\alpha}^{\beta\rho}u^\alpha - 2Q^{\rho\alpha\beta}u_\alpha  \right)\nabla_\beta u_\rho - \frac{1}{\ell^2}A^\rho a_\rho - \nabla_\rho a^\rho - 2{Q^{\rho\alpha}}_\rho a_\alpha - \frac{1}{\ell^2}u^\rho \Df_u a_\rho \\+ \frac{1}{n-1}\theta^2 - \omega^2 + \sigma^2 + X^2 = -R_{\alpha\beta} u^\alpha u^\beta + 2 T^{\rho\alpha\beta} u_\alpha \nabla_\rho u_\beta + 4 T^{\rho\alpha\beta} Q_{\rho\alpha\epsilon}u_\beta u^\epsilon
	\end{multline}
	Rearranging the above equation, we finally get the Raychaudhuri equation in a familiar form,
	\begin{equation}\label{rayeq}
		\Df_u \theta + \frac{1}{n-1}\theta^2 =  -R_{\alpha\beta} u^\alpha u^\beta + \omega^2 - \sigma^2 + \nabla_\rho a^\rho + \mathcal{N}
	\end{equation}
	where, we have combined the extra terms coming from the non-Riemannian nature of the geometry as a single scalar, $\mathcal{N}$, given by,
	\begin{equation}
		\mathcal{N} := \left(2 T^{\beta\alpha\rho} + 4Q^{[\rho\alpha]\beta}  \right)u_\alpha\nabla_\beta u_\rho + 2{Q^{\rho\alpha}}_\rho a_\alpha + 4 T^{\rho\alpha\beta} Q_{\rho\alpha\epsilon}u_\beta u^\epsilon + \frac{1}{\ell^2} \Df_u \left(u^\rho a_\rho\right) - X^2
	\end{equation}
	Equation \eqref{rayeq} is the Raychaudhuri equation in a general non-Riemannian geometry. In the Riemannian limit, we have, $\boldsymbol{Q} = 0 = \boldsymbol{T}$, and $u^\rho a_\rho = 0 \Rightarrow X^2 = 0$, which makes $\mathcal{N} = 0$.

	\section{Lagrangian Formulation} \label{sec-raylagham}
	To construct a Lagrangian for the RE, it is useful to first convert it in into a second-order differential equation. In Riemannian geometry, this is done by recognising that the expansion scalar, $\theta$, is equal to the fractional rate of change of the cross-sectional volume of the congruence. A formal proof of this can be found in \cite{poisson}. Then, recasting the RE in terms of the volume element of the congruence's cross-section, one can construct a Lagrangian using this volume element as the dynamical degree of freedom. Therefore, in order to construct a Lagrangian for the non-Riemannian RE, we will first follow a similar procedure as in \cite{poisson} to establish a relationship between the expansion scalar and the cross-sectional volume. 
	
	\subsection{Expansion Scalar and Cross-Sectional Volume}
	To introduce the notion of a cross-sectional volume, we select a particular point, $ P $, corresponding to a parameter value, $ t_P $, on any given curve in the congruence. Then, the cross-section around this curve at point, $ P $, can be constructed by taking a set of points, $ P^\prime $, on the neighbouring curves which have the same parameter value, $ t_P $. Let us call this set of points $\Sigma_P$. To construct a metric on $ \Sigma_P $, we introduce a coordinate system, $ y^a\ (a = 1,2,\dots, n-1) $. The vectors, $ n^\alpha_a = \frac{\partial x^\alpha}{\partial y^a} $, are tangent to the cross-section. In other words, $n^\alpha_a$ is orthogonal to the congruence, i.e., $n^\alpha_a u_\alpha = 0$. Now, for a distance between two points on the cross-section, we have $ \df t = 0 $, and hence, we can write,
	\begin{align}
		\df s^2 &= \cg_{\alpha\beta}\df x^\alpha \df x^\beta = \cg_{\alpha\beta} \frac{\partial x^\alpha}{\partial y^a}\df y^a \frac{\partial x^\beta}{\partial y^b}\df y^b \\
		\Rightarrow \df s^2 &= \cg_{\alpha\beta} n^\alpha_a n^\beta_b \df y^a \df y^b = h_{a b} \df y^a \df y^b
	\end{align}
	Therefore, the $ (n-1) $-tensor, $ h_{ab} = \cg_{\alpha\beta} n^\alpha_a n^\beta_b $, acts like the $ (n-1) $-metric of the cross-section. The volume element on the cross-section would then be given by, $ \delta V = \sqrt{h}\ \df^{n-1} y$, where, $h = \det h_{ab}$. Since $n^\alpha_a u_\alpha = 0$, we can also write, $ h_{ab}~=~h_{\alpha\beta} n^\alpha_a n^\beta_b $.
	
	To calculate a rate of change of volume, we want to compare $ \Sigma_P $ to another such set of points, say, $ \Sigma_Q $, at some other point, $ Q $, on the curve where the parameter value is $ t_Q $. Since each point, $ P^\prime $, on the cross section is also a point on a neighbouring curve, the coordinate system, $y^a$, can also be used as a label for these neighbouring curves. Then, identifying all the points on $ \Sigma_Q $ by the curves passing through them, we automatically get a coordinate system on $ \Sigma_Q $ as well. Since we set up the coordinates, $ y^a $, such that they remain the same along the curves (that is a curve retains its label), the change in the volume element as one goes from $ P $ to $ Q $ is only due to the change in $ \sqrt{h} $. Therefore, one can write,
	\[ \frac{1}{\delta V}\Df_u \delta V  = \frac{1}{\sqrt{h}}\Df_u \sqrt{h} = \frac{1}{ 2h_{ab}}\Df_u h_{ab} \]
	
	To calculate the fractional rate of change of the volume, we start with,
	\begin{align}
		\Df_u h_{a b} &= \Df_u (\cg_{\alpha\beta}n^\alpha_a n^\beta_b) = \left(2P_{\rho\alpha\beta}u^\rho + 2\nabla_{(\alpha}u_{\beta)} \right)n^\alpha_a n^\beta_b \nonumber \\
		\Rightarrow \frac{1}{ 2h_{ab}}\Df_u h_{ab} &= \left(P_{\rho\alpha\beta}u^\rho + \nabla_{(\alpha}u_{\beta)} \right)h^{\alpha\beta} \nonumber\\
		\Rightarrow \frac{1}{\delta V}\Df_u \delta V &= \theta + {P_{\rho\alpha}}^\alpha u^\rho + \frac{1}{\ell^2}P_{\rho\alpha\beta}u^\rho u^\alpha u^\beta = \theta + \phi
	\end{align}
	where, $\phi = {P_{\rho\alpha}}^\alpha u^\rho + \frac{1}{\ell^2}P_{\rho\alpha\beta}u^\rho u^\alpha u^\beta$.
	
	Therefore, instead of being exactly equal to it (as in the Riemannian case), the expansion scalar here is related to the fractional rate of change of the cross sectional $ (n-1) $-volume of the congruence through a scalar, $ \phi $, that depends on the non-metricity. There are three points to be noted here: (i) due to the scalar, $\phi$, one cannot use the volume element directly as the dynamical dof in formulating a Lagrangian for RE; (ii) since $\phi$ depends only on the projection of non-metricity along the congruence, there exists a subclass of non-Riemannian geometries (with $P_{\rho\alpha\beta}u^\rho=0$) for which the volume element does act as a dynamical dof; and (iii) for non-Riemannian geometries with only torsion (i.e., $\boldsymbol{Q} = 0$), the volume element again acts as the dynamical variable.
	
	\subsubsection{A Case when $\phi = 0$}
	The directional derivative of the $n$-velocity is given by,
	\begin{equation}
		A^\alpha := u^\beta\nabla_\beta u^\alpha \;\; \Rightarrow\ 2u_\alpha A^\alpha = u^\beta \nabla_\beta(-\ell^2) + P_{\rho\alpha\beta}u^\rho u^\alpha u^\beta
	\end{equation}
	Therefore, if one considers fixed length curves that are autoparallel with respect to the non-Riemannian connection $(\boldsymbol{A} = 0)$ (or curves whose path-acceleration ($\boldsymbol{A}$) is orthogonal to the velocity), we get, $P_{\rho\alpha\beta}u^\rho u^\alpha u^\beta = 0 $. Further, for a Weyl type non-metricity \cite{weyl1,hehl1} ($Q_{\rho\alpha\beta} = {Q_{\rho\epsilon}}^\epsilon \cg_{\alpha\beta}$), this would mean that ${P_{\rho\alpha}}^\alpha u^\rho = 0$. Combined together, this results in $\phi = 0$. We present this case only as a quick example. There might exist other forms of the non-metricity tensor or particular types of congruences as well, for which $\phi$ vanishes.
	
	\subsection{Constructing a Lagrangian}
	We started with establishing a relation between the cross-sectional volume element and the expansion scalar with the hope that we would be able to convert the RE into a second-order differential equation. However, that does not seem to be as straightforward as in Riemannian geometry. To reconcile this problem, we introduce another scalar, say, $\psi$, such that, $\Df_u(\ln \psi) = \phi$. Using this, the equation for the fractional rate of change of cross-sectional volume becomes,
	\begin{equation}
		\Df_u (\ln \delta V) = \theta + \Df_u(\ln \psi) \Rightarrow \Df_u\left[\ln\left(\frac{\delta V}{\psi}\right)\right] = \theta
	\end{equation}
	Defining a new variable, $q = \frac{\delta V}{\psi}$, the above equation becomes,
	\begin{equation} \label{qdef}
		\frac{1}{q}\Df_u q = \theta
	\end{equation}
	Therefore, the expansion scalar in a non-Riemannian geometry can be thought of as being equal to the fractional rate of change of the volume element with a `weight' given by a scalar factor that depends on the non-metricity. It is this quantity, $q = \psi^{-1}\delta V$, that we will consider as our dynamical degree of freedom. Using Eq. \eqref{qdef} in Eq. \eqref{rayeq}, we get,
	\begin{equation} \label{rayeq2}
		\frac{1}{q}\Df^2_u q - m\frac{1}{q^2}\left(\Df_u q\right)^2 =  -R_{\alpha\beta} u^\alpha u^\beta + \omega^2 - \sigma^2 + \nabla_\rho a^\rho + \mathcal{N}
	\end{equation}
	where, $m = \frac{n-2}{n-1}$. The above equation is a second-order differential equation which can be derived as the dynamical equation of a Lagrangian. Based on the analysis in \cite{chakra}, one can construct the following Lagrangian,
	\begin{equation}
		\mathcal{L} = \hf q^{2 m} \left(\Df_u q\right)^2 - U(q)
	\end{equation}
	In analogy with a mechanical system, the first term can be considered as a `kinetic' term (square of first differential) and second term as a `potential'. The corresponding Euler-Lagrange equations are given by,
	\begin{align}
		\frac{\partial \mathcal{L}}{\partial q} - \Df_u \left[\frac{\partial \mathcal{L}}{\partial\left(\Df_u q\right)}\right] &= 0 \\
		\Rightarrow \frac{1}{q}\Df^2_u q - m\frac{1}{q^2}\left(\Df_u q\right)^2 &= \frac{\partial U}{\partial q} q^{-(2m+1)}
	\end{align}
	The above equation matches the RE (Eq. \eqref{rayeq2}) with the potential defined in the following manner,
	\begin{equation}\label{potdef}
		\frac{\partial U}{\partial q} = \left( -R_{\alpha\beta} u^\alpha u^\beta + \omega^2 - \sigma^2 + \nabla_\rho a^\rho + \mathcal{N}\right) q^{2m+1}
	\end{equation}
	
	\subsection{Hamiltonian Formulation and Poisson Brackets}
	Once we have the Lagrangian for the RE, constructing the Hamiltonian becomes straightforward. Using the canonical definition, the Hamiltonian can be defined as,
	\begin{equation}
		\mathcal{H} = \frac{\partial \mathcal{L}}{\partial\left(\Df_u q\right)} \Df_uq - \mathcal{L} = \hf q^{2 m} \left(\Df_u q\right)^2 + U(q)
	\end{equation}
	Then, the first Hamilton's equation gives the conjugate momentum, $p = q^{2m}\Df_u q $. The second one is given by,
	\begin{align}
		\Df_u p &= - \frac{\partial \mathcal{H}}{\partial q} \\
		\Rightarrow \frac{1}{q}\Df^2_u q - m\frac{1}{q^2}\left(\Df_u q\right)^2 &= \frac{\partial U}{\partial q} q^{-(2m+1)}
	\end{align}
	which is the same as the Raychaudhuri Eq. \eqref{rayeq2} using the definition of the potential \eqref{potdef}.
	
	Using the above definition of the Hamiltonian, it is easy to see that the RE can also be written in terms of Poisson brackets as,
	\begin{equation}
		\{p,\mathcal{H}\} = 0
	\end{equation}
	
	
	\section{Discussion} \label{sec-discuss}
	In this paper, we generalised the Raychaudhuri equation to non-Riemannian geometry and then presented a Lagrangian formulation for it. Treating the Raychaudhuri equation as the dynamical equation for the expansion of curves, we first recognised an appropriate dynamical degree of freedom that can be used to construct a Lagrangian. In Riemannian geometry, this is simply the volume element of the congruence's cross-section. However, in non-Riemannian geometry, we found that this dof is given by a `weighted' volume element with the weight given by, $\psi^{-1} = \e^{-\int \left({P_{\rho\alpha}}^\alpha u^\rho + \frac{1}{\ell^2}P_{\rho\alpha\beta}u^\rho u^\alpha u^\beta\right)\df t }$. Treating the congruence of timelike curves as a dynamical system with this weighted volume element as the dynamical degree of freedom, we constructed a Lagrangian such that the corresponding Euler-Lagrange equation gave us the Raychaudhuri equation. From this Lagrangian, we constructed a Hamiltonian and showed that the one of the Hamilton's equations corresponds to the Raychaudhuri equation. Further, Poisson brackets for the RE were also presented.
	
	The Raychaudhuri equation is a geometric evolution equation for the trace part of the velocity gradient (also called expansion scalar) of a congruence of timelike curves. In GR, it plays an instrumental role in cosmology, studies of gravitational collapse, singularity theorems, black hole mechanics, etc. \cite{kar}. The Lagrangian formulation of it has been used in studies of geometric flows \cite{alsa,horw}. The results in this paper will be useful in studying these issues in the context of non-Riemannian theories of gravity. They may also prove useful in studies of quantum gravity (e.g., for the analysis of (quantum) singularity theorems) within these theories (see \cite{das,chakra,alsa} and also \cite{lash,das2}). Further, in \cite{horw}, a connection was established between the expansion scalar and the geometric entropy. It would be a worthwhile to check if this result can be generalised to non-Riemannian geometries.
	
	It would also be of interest to extend this analysis to the case of null curves. However, the non-constancy of the vector norms would mean that a null curve may not remain null under parallel transportation. However, particular forms of non-metricity that allow for fixed length vectors can be used after which, one would essentially follow the same steps as presented in this paper. Finally, it is important to note that the Raychaudhuri equation only describes the evolution of the expansion of the congruence. Therefore, to characterise the congruence (flow) completely and one would also need a Lagrangian formulation for the other kinematic quantities (rotation and shear). 
	
	\section*{Acknowledgements}
	We thank Sai Madhav Modumudi for useful discussions and comments.

	\bibliographystyle{unsrtnat}	
	\bibliography{raylag-arxiv}
	
\end{document}